\documentclass[RNAAS]{aastex631}

\begin{document}

\title{The JWST Resolved Stellar Populations Early Release Science Program I.: NIRCam Flux Calibration}

\author[0000-0003-4850-9589]{Martha L.\ Boyer}
\affiliation{Space Telescope Science Institute, 3700 San Martin Drive, Baltimore, MD 21218, USA}


\author[0000-0003-2861-3995]{Jay Anderson}
\affiliation{Space Telescope Science Institute, 3700 San Martin Drive, Baltimore, MD 21218, USA}

\author[0000-0002-5581-2896]{Mario Gennaro}
\affiliation{Space Telescope Science Institute, 3700 San Martin Drive, Baltimore, MD 21218, USA}

\author[0000-0002-7007-9725]{Marla Geha}
\affiliation{Department of Astronomy, Yale University, New Haven, CT 06520, USA}

\author[0000-0001-5538-2614]{Kristen B.\ Wingfield McQuinn}
\affiliation{Rutgers University, Department of Physics and Astronomy, 136 Frelinghuysen Road, Piscataway, NJ 08854, USA}

\author[0000-0002-9599-310X]{Erik Tollerud}
\affiliation{Space Telescope Science Institute, 3700 San Martin Drive, Baltimore, MD 21218, USA}

\author[0000-0001-6464-3257]{Matteo Correnti}
\affiliation{Space Telescope Science Institute, 3700 San Martin Drive, Baltimore, MD 21218, USA}

\author[0000-0002-8092-2077]{Max J. Brenner Newman}
\affiliation{Rutgers University, Department of Physics and Astronomy, 136 Frelinghuysen Road, Piscataway, NJ 08854, USA}

\author[0000-0002-2970-7435]{Roger E. Cohen}
\affiliation{Rutgers University, Department of Physics and Astronomy, 136 Frelinghuysen Road, Piscataway, NJ 08854, USA}

\author[0000-0002-3204-1742]{Nitya Kallivayalil}
\affiliation{Department of Astronomy, University of Virginia, 530 McCormick Road, Charlottesville, VA 22904, USA}

\author[0000-0002-1691-8217]{Rachel Beaton}
\affiliation{Space Telescope Science Institute, 3700 San Martin Drive, Baltimore, MD 21218, USA}
\affiliation{Department of Astrophysical Sciences, Princeton University, 4 Ivy Lane, Princeton, NJ 08544, USA}
\affiliation{The Observatories of the Carnegie Institution for Science, 813 Santa Barbara St., Pasadena, CA 91101, USA}

\author[0000-0003-0303-3855]{Andrew A. Cole}
\affiliation{School of Natural Sciences, University of Tasmania, Private Bag 37, Hobart, Tasmania, 7001 Australia}

\author[0000-0001-8416-4093]{Andrew Dolphin}
\affiliation{Raytheon, Tucson, AZ 85726, USA; Steward Observatory, University of Arizona, Tucson, AZ 85726, USA}

\author[0000-0001-9690-4159]{Jason S. Kalirai}
\affiliation{Johns Hopkins University Applied Physics Laboratory, 11100 Johns Hopkins Road, Laurel, MD 20723, USA}

\author[0000-0002-4378-8534]{Karin M. Sandstrom}
\affiliation{Center for Astrophysics and Space Sciences (CASS), University of California San Diego, 9500 Gilman Drive, La Jolla, CA 92093, USA}

\author[0000-0002-1445-4877]{Alessandro Savino}
\affiliation{Department of Astronomy, University of California, Berkeley, Berkeley, CA 94720, USA}

\author[0000-0003-0605-8732]{Evan D. Skillman}
\affiliation{Minnesota Institute for Astrophysics, University of Minnesota, 116 Church Street SE, Minneapolis, MN 55455, USA}

\author[0000-0002-6442-6030]{Daniel R.\ Weisz}
\affiliation{Department of Astronomy, University of California, Berkeley, Berkeley, CA 94720, USA}

\author[0000-0002-7502-0597]{Benjamin F.\ Williams}
\affiliation{Department of Astronomy, Box 351580, University of Washington, Seattle, WA 98195, USA}

\begin{abstract}
    We use globular cluster data from the Resolved Stellar Populations Early Release Science (ERS) program to validate the flux calibration for the Near Infrared Camera (NIRCam) on the James Webb Space Telescope ({\it JWST}). We find a significant flux offset between the eight short wavelength detectors, ranging from 1--23\% ($\sim$0.01-0.2~mag) that affects all NIRCam imaging observations. We deliver improved zeropoints for the ERS filters and show that alternate zeropoints derived by the community also improve the calibration significantly. We also find that the detector offsets appear to be time variable by up to at least 0.1~mag.
\end{abstract}


\section{Introduction}

One of the goals of the {\it JWST} Resolved Stellar Populations ERS program (PID 1334; D. Weisz et al., in preparation) is to optimize the photometric performance of the NIRCam instrument \citep{Rieke+2005}. In this note, we use the ERS observations of globular cluster M92 (NGC 6341) to illustrate the current status of NIRCam's relative flux calibration. As one of the lowest metallicity Galactic clusters \citep[$\mathrm{[Fe/H]} = -2.35\pm0.05$;][]{carretta09}, M92 is often used as a calibrator for extragalactic stellar population studies \citep{Brown2005}.




NIRCam consists of 8 separate shortwave (SW) detectors split into two modules (A1-A4 and B1-B4) and 2 longwave (LW) detectors.   Using the M92 stellar main sequence measured with NIRCam, we find a significant relative flux offset between the eight SW detectors when using the zeropoints in the latest pipeline reference files (jwst\_0959.pmap),\footnote{\href{https://jwst-crds.stsci.edu/}{https://jwst-crds.stsci.edu/}} which has also been noticed by other teams.\footnote{\label{gabe}See G.\ Brammer's zeropoint corrections on Github \href{https://github.com/gbrammer/grizli/pull/107}{https://github.com/gbrammer/grizli/pull/107}}  We also investigate the time variability of these offsets by analyzing several epochs of LMC data in the {\it JWST} archive.

\subsection{NIRCam Absolute Flux Calibration}

The current NIRCam absolute flux zeropoints were derived with Commissioning data (PID 1074) that imaged the G-type standard stars P330E and P177E in all filters using subarrays on SW detector B1 and LW detector BLONG. The measured fluxes were then translated to the remaining 8 detectors by overlapping all detectors on the same set of stars in the LMC Calibration field \citep{vanderMarel+2007, Sahlmann+2019}. Only a subset of filters were used in the LMC observations; zeropoints for the remaining filters were derived by interpolating the standard star spectra.  

{\it JWST} Commissioning was intended to provide only a first pass at the absolute flux calibration. A Cycle 1 Calibration program is underway to deliver more robust measurements that include additional flux standards imaged in every filter on each of the 10 detectors.  This will forgo the need to use the LMC to transfer the calibration between detectors \citep{Gordon+2022}.  In the meantime, we can test the zeropoints in  the current set of pipeline reference files using the M92 ERS data.

\section{M92 Observations and Photometry}

The ERS NIRCam observations centered M92 in the $\approx$43\arcsec\ gap between modules A and B, capturing the east and west sides of the cluster while avoiding the crowded and saturated center.  We selected the F090W and F150W filters in the SW channel and the F277W and F444W filters in the LW channels. The total exposure time was 1245s per filter, achieved with the SHALLOW4 readout pattern, 6 groups, and 4 subpixel dithers.\footnote{DOI: \dataset[10.17909/hdf2-je40]{http://dx.doi.org/10.17909/hdf2-je40}} We reduced the data with the latest available reference files (jwst\_0959.pmap)\footnote{\href{https://jwst-crds.stsci.edu/display_context_history/}{https://jwst-crds.stsci.edu/display\_context\_history/} Newer pmaps are available (2022 Aug 31), but they do not include NIRCam imaging updates.}, with zeropoints that were delivered to the {\it JWST} Calibration Reference Data System (CRDS) and incorporated into the pipeline on 28 July, 2022.


The HST1PASS tool \citep{HST1PASS} was designed to perform point spread function (PSF) photometry on Hubble Space Telescope ({\it HST}) images. This tool finds stars and performs photometry on the individual exposures using an empirical ``effective" PSF model. A version of HST1PASS for use on {\it JWST} data is under development, and we use an early version here (JWST1PASS; J.\ Anderson, private communication).

\begin{figure}[ht!]
\plotone{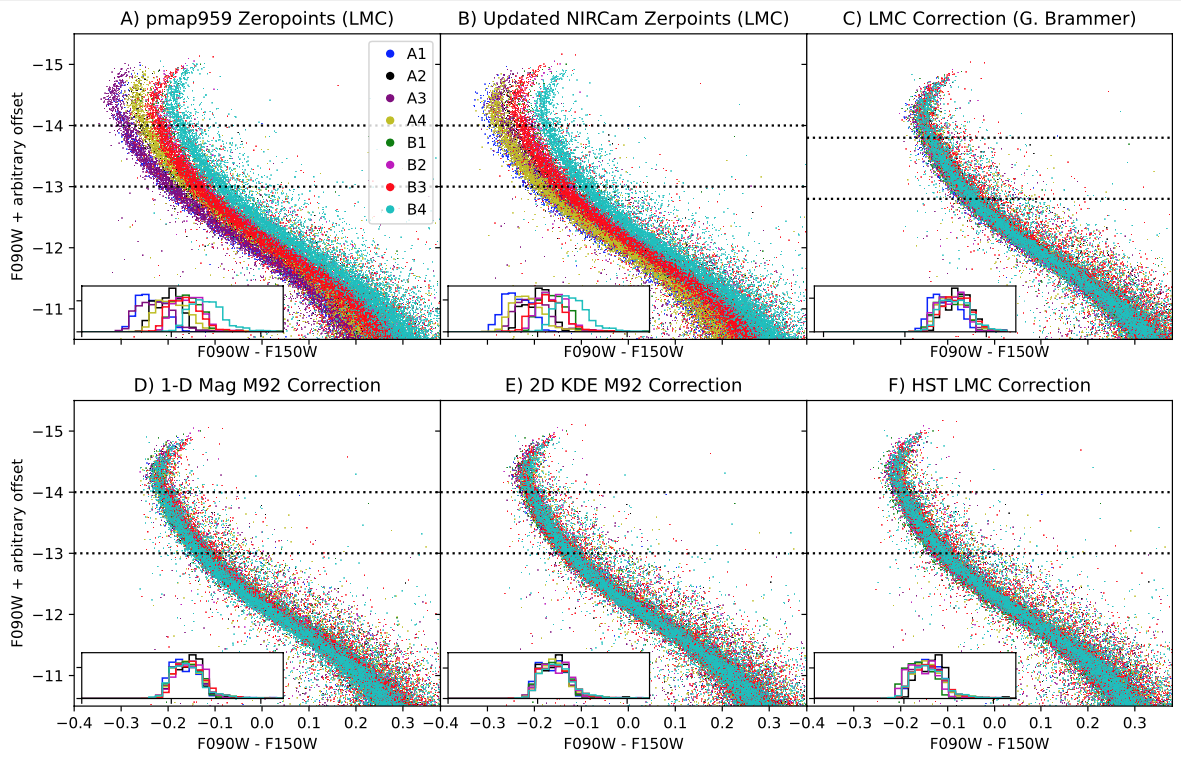}
\caption{M92 CMDs,  plotted with different zeropoints and color-coded by NIRCam SW detectors. Inset in each panel is a histogram of the F090W--F150W color between the dotted lines, offset in panel C to account for the different absolute calibration. The zeropoints used to make these CMDs are available in a table behind the figure. {\it Panel A:} Zeropoints currently in the {\it JWST} pipeline (pmap959), derived with Commissioning data (PID 1074). Several distinct (and overlapping) branches are visible, caused by detector flux offsets. {\it Panel B:} Plotted using the latest NIRCam zeropoints, not yet in the {\it JWST} pipeline.  {\it Panel C:} Zeropoints from G. Brammer, derived from LMC data and bootstrapped to the NIRISS flux calibration.  {\it Panel D:} 1-D luminosity function correction derived from M92 data, normalized to detector B1. {\it Panel E:} 2-D KDE correction derived from M92 data normalized to detector B1. {\it Panel F:} Offsets derived from LMC data in the archive, normalized to detector B1.
\label{fig:cmd}}
\end{figure}

Panel A of Figure~\ref{fig:cmd} shows the resulting color-magnitude diagram (CMD) for M92 in the SW filters. The main sequence is divided into several overlapping, but distinct branches that coincide with different SW detectors, indicating offsets in the flux calibration between detectors that are clearly illustrated by this exquisite dataset. The inset at the bottom of the panel shows the color histogram for $\approx$230 stars, about a magnitude below the main sequence turnoff (outlined by dotted lines), that clearly show the color differences between the detectors. 

The NIRCam team recently derived new absolute flux zeropoints from the Commissioning data (M. Rieke, private communication), this time using the latest flat fields delivered to CRDS on 2022 Aug 19. The CMD resulting from these zeropoints is shown in Panel B, showing some improvement, but with a clear detector offset remaining.  These new zeropoints have not been ingested into the {\it JWST} pipeline as of publication, but they are available in the table behind Figure~\ref{fig:cmd}.

In Panel C of Fig.~\ref{fig:cmd}, we show zeropoints derived by G. Brammer, and available on GitHub$^2$. That team used LMC data in the {\it JWST} archive to derive detector offsets, and tied the absolute calibrations to those measured for the NIRISS detector.

\subsection{Deriving Flux Offsets Between Detectors}

We derive our own flux offsets between the SW detectors using 3 techniques:   2 based on M92 and 1 on LMC data.
First, we adjust the M92 luminosity functions in F090W and F150W for each detector to match the luminosity function of B1, since the standard stars were observed directly on B1. The resulting CMD is shown in Panel D of Figure~\ref{fig:cmd}.

As an alternative approach, we fit the detector offsets using a kernel density estimator (KDE) with F090W and F150W magnitudes as the two KDE dimensions.\footnote{scipy.stats.gaussian\_kde}  We then use a Levenberg-Marquardt optimizer to find the per detector zeropoints that minimize the residuals in the KDE between that detector and the B1 KDE, using the sources in B1 as the ``grid" over which to compute the KDE. The resulting CMD is shown in Panel E of Figure~\ref{fig:cmd}.


Finally, we use all available LMC observations to derive offsets: Commissioning PIDs 1069, 1072, 1073, 1074, 1473, and Calibration PID 1476, processed with the same reference files used for the M92 reductions (jwst\_0959.pmap). We perform aperture photometry on each LMC image and match the resulting catalogs to the {\it HST} F606W catalog \citep{vanderMarel+2007}, which is accurate to the 1\% level.  Detector offsets are measured by comparing the color of the main sequence ridge line for each detector, in the F606W--F090W and F606W--F150W colors. We show the resulting M92 CMD in Panel F of Figure~\ref{fig:cmd}. Similar analysis on additional filters is forthcoming.


After applying each set of offsets, the combined M92 main sequence widths vary by $<$0.04~mag along the magnitude range shown here.  However, the 2D KDE technique results in the best total overlap between detectors, which is illustrated by the histograms inset in each panel.  Note, the $\approx0.2$~mag difference in brightness in panel C is due to the different choice in the {\it absolute} calibration: we bootstrap our offsets in Panels D, E, and F to detector B1, which was used to measure the flux standard stars, while the Brammer offsets bootstrap to the NIRISS absolute flux calibration.

\subsection{Zeropoint Time Variability}


The six LMC programs we analyzed to derive the offsets shown in Panel F were observed between 2022 Apr 04 and Jul 11, with F150W in every epoch.  Examination of each epoch individually reveals clear time variability in the F150W zeropoints. The offsets between epochs are significant, with a range of  $\sim$0.02--0.1~mag.  The photometric stability will be monitored more closely by the Cycle 1 Absolute Flux program, which is observing the same standard star each month \citep{Gordon+2022}.

\section{Conclusions}

The ERS data of M92 clearly demonstrate a significant flux offset between the eight SW NIRCam detectors in the current {\it JWST} pipeline, on the order of 1-23\% ($\sim$0.01--0.2~mag). An absolute flux calibration program is expected to improve these measurements throughout Cycle 1 \citep{Gordon+2022}. In the meantime, the results from our 2-D KDE correction provide the best offsets for F090W and F150W. Zeropoints derived for additional filters by the community (G. Brammer$^2$) provide good results in these two filters, though there is room for improvement on the order of $<$0.04~mag. We also discover significant offsets (up to $\approx$0.1~mag) in observations of the LMC that span about 3 months. 



\newpage

\begin{acknowledgements}
We thank Marcia Rieke for providing the latest zeropoints derived from the NIRCam Commissioning program. This work is based on observations made with the NASA/ESA/CSA James Webb Space Telescope. The data were obtained from the Mikulski Archive for Space Telescopes at the Space Telescope Science Institute, which is operated by the Association of Universities for Research in Astronomy, Inc., under NASA contract NAS 5-03127 for {\it JWST}. These observations are associated with programs 1334, 1473, 1476, 1069, 1072, 1073, and 1074. We acknowledge the NIRCam team for developing calibration program 1476 with a zero-exclusive-access period.
\end{acknowledgements}

\vspace{5mm}
\facilities{JWST (NIRCam), MAST}

\software{astropy \citep{2013A&A...558A..33A,2018AJ....156..123A},
         scikit-learn \citep{scikit-learn}, scipy \citep{SciPy2020} }

\newpage

\bibliography{main}{}
\bibliographystyle{aasjournal}

\end{document}